\documentstyle[aps]{revtex}

\begin{document}

\title{{\Large {\bf Exotic Acceleration
Processes and Fundamental Physics}}}

\author{{\bf Giovanni~AMELINO-CAMELIA}}
\address{Dipartimento di Fisica, Universit\`{a} di Roma ``La Sapienza''
and INFN Sez.~Roma1,\\
    P.le Moro 2, 00185 Roma, Italy}

\maketitle

\begin{abstract}\noindent%
Gamma-ray bursts and ultra-high-energy cosmic rays provide
an important testing ground for fundamental physics.
A simple-minded analysis of some gamma-ray bursts would
lead to a huge estimate of the overall energy emitted,
and this represents a potential challenge for modelling the bursts.
Some cosmic rays have been
observed with extremely high energies, and it is not easy to
envision mechanisms for the acceleration of particles to such high
energies.  Surprisingly some other aspects
of the analysis of gamma-ray bursts and ultra-high-energy cosmic
rays, even before reaching a full understanding of the mechanisms
that generate them, can already be used to explore new ideas in
fundamental physics, particularly for what concerns the structure
of spacetime at short (Planckian) distance scales.
\end{abstract}

%
%

%


%
%

\section{Introduction}
Much of the work in astrophysics concerns the use of
the presently-accepted laws of fundamental physics
in the development of consistent descriptions of the observations.
In some cases the key challenge comes from an energy-balance issue.
For example, for some gamma-ray bursts the overall observed fluences
can be as large as $10^{-4}ergs/cm^2$, which would correspond to
a luminosity of $10^{52}ergs/s$ and higher, if one can assume
isotropy. It would be difficult to describe such levels of luminosity
in terms of conventional physics.
We now understand~\cite{tsviGRBnewREV} that most gamma-ray bursts
are narrowly beamed (not at all isotropic) and this lowers the luminosity
estimate to a more mundane level; however, there is still
no real consensus on the mechanism of emission
of gamma-ray bursts.

Another energy-balance issue is encountered in the study of cosmic rays.
In fact, recent data suggest~\cite{agasa}~\cite{fly} that
some cosmic rays have energies of $10^{20} eV$ and higher.
And it has been argued~\cite{bwCR}
that at least some of these cosmic rays could originate
from within our galaxy. But we are unable to find within our galaxy
sources which are good candidates for accelerating particles
to such high energies.

In the study of these issues some ``exotic acceleration
mechanisms'' are being considered.
Usually the relevant acceleration mechanisms
are deemed ``exotic'' not because of some role for
new laws of physics, but rather because
of the role played by creative ways
to rely on the presently-accepted laws of fundamental physics
to devise mechanisms that could explain the energetics of the
system of interest.
Only in a few of the most speculative papers on these subjects
it has been argued that new physics might be responsible
for the emission mechanism.

While the possibility of new physics is at best a marginal hypothesis
for the study of emission mechanisms, surprisingly these fascinating
problems in astrophysics do provide, independently of the understanding
of the emission mechanisms, some of the best arenas for testing ideas
for new laws of fundamental physics, especially for what concerns
the structure of spacetime at Planckian distance scales.
We do not need to know exactly how gamma-ray bursts are emitted
in order to observe that a gamma-ray burst is a very rich signal
which propagates over very large distances. The analysis of gamma-ray bursts
is therefore a wonderful opportunity for high-sensitivity studies of
the laws of propagation of signals through space.
Similarly we do not need to know exactly how a $10^{20}eV$ cosmic ray
is produced in order to observe that the collisions
between such a cosmic ray and the low-energy photons it encounters
on the way to our Earth laboratories provide a rare opportunity
to test the nature of boost transformations.
In fact, the same collisions are studied in our
particle-physics laboratories but only relatively
close to the center-of-mass frame.

\section{On the emission of gamma-ray bursts}
Even taking into account
the fact that most gamma-ray bursts
are narrowly beamed,
in some cases the inferred luminosity is as high
as $10^{51}ergs/s$.
This is still rather large
but does  not represent a record-setting energy release,
since the total energy emitted is
comparable to the one of supernovae~\cite{tsviGRBnewREV}.

The structure of a gamma-ray burst, viewed from a signal-analysis
perspective, is extremely rich, and this of course represents
a challenge for emission models.
We need a model that would be able to reproduce
faithfully all aspects of that rich structure.
A popular idea is the one of the ``fireball internal-external
shocks model''~\cite{tsviGRBnewREV}, but it is perhaps too early
to speak of a general consensus.

An example of the issues that are being discussed in the effort
of establishing the emission mechanism is provided by
the contribution by Kaneko {\it et al}~\cite{kaneko},
which stresses how
important insight on the emission mechanism of gamma-ray bursts
can be obtained by establishing the proper description
of gamma-ray-burst spectra.
On the basis of an interesting broadband spectral analysis
of two spectrally-hard gamma-ray bursts Kaneko {\it et al}
conclude that gamma-ray-burst spectra evolve on time scales
that are much longer than the synchrotron cooling time.
This may be consistent with
some arguments
discussed in the gamma-ray-burst literature
(see, {\it e.g.}, Ref.~\cite{tsviGRBREV}),
but Kaneko {\it et al}
argue that instead this should lead to the
conclusion that the acceleration
mechanism is more complicated than predicted
by some popular gamma-ray-burst models.

\section{Gamma-ray bursts as a quantum-spacetime laboratory}
Gamma-ray bursts have recently attracted much
interest (see, {\it e.g.}, the recent
reviews~\cite{polonpap}~\cite{gacqm100}~\cite{qgpSubir}~\cite{qgpNick})
also from the research community exploring the
hypothesis that spacetime might have to be ``quantized'',
{\it i.e.} that there might be a quantization of
spacetime observables
in a way somehow analogous to the quantizations of all
other observables
encountered in other aspects of fundamental physics.

Some scenarios for
spacetime quantization
involve a sort of granularity of spacetime,
and as a result one may expect some departures
from the smooth laws of Lorentz symmetry~\cite{grbgac}.
The mechanism would be analogous\footnote{Some authors have
also argued (see, {\it e.g.}, Ref.~\cite{garayPRL}) that
the quantum-spacetime environment might act in a way that
to some extent resembles the one of  a thermal environment.
It is well established (see, {\it e.g.}, Ref.~\cite{gacpi})
that in a thermal environment the energy-momentum dispersion relations
are naturally modified.} to the one
that applies to phonons: the law of propagation of phonons
is formally relativistic
at low energies, when the reticular structure of the
underlying material can be ignored, but at high energies
there are departures from the relativistic behaviour.
If spacetime itself was granular then some analogous effect
might be present. A sort of light-cone fuzziness is essentially
inevitable in presence of spacetime granularity, and recently it
was realized that some specific schemes for introducing the
granularity length scale may also affect
the propagation of photons by introducing a small dependence of the speed
on the photon energy.

These effects on photon propagation are expected
to be very small,
since their magnitude is set by the ratio
of the photon energy over the Planck energy scale ($\sim 10^{28} eV$),
but, as I stress in my contribution~\cite{gachunt},
through the analysis of observations of gamma-ray bursts one has a chance
to discover (or rule out) these Planck-scale effects.
The properties of gamma-ray bursts that are used in this type
of analysis are~\cite{grbgac}~\cite{billetal}

$\bullet$ the fact that gamma-ray bursters are often
at cosmological distances,

$\bullet$ the fact that a typical gamma-ray-burst spectrum
should extend up to the tens of $MeV$ and higher,

$\bullet$ and the fact that some ``microbursts''
within a gamma-ray burst can have very short duration,
as short as $10^{-3} s$ (or even $10^{-4} s$).

The key point I want to stress here is that
this use of gamma-ray burst for quantum-spacetime studies
is largely insensitive on the nature of the emission mechanism.
The properties of gamma-ray bursts which are used
are well established, and the analysis is largely independent
of the modelling of the emission mechanism.
So the convergence of interest on gamma-ray bursts
form the exotic-acceleration-mechanisms community
and the quantum-spacetime community is accidental.

\section{On the most energetic cosmic rays}
Several mysteries surround the observation of cosmic rays.
This is particularly true for the ``ultra-high-energy (UHE) cosmic rays'',
with energies higher than $10^{19}eV$.
The identification of the particles is problematic, since we reveal
them indirectly through their interactions in the atmosphere.
It appears however that most UHE cosmic rays are protons.

Most UHE cosmic rays are believed to be of cosmological origin,
but this then should imply that the so-called ``GZK
cutoff''~\cite{gzk} should be observed: the spectrum of observed
cosmic rays should basically stop around $E_{gzk} \simeq 5 \cdot
10^{19}eV$, where photons in the Cosmic Microwave Background (CMB)
become viable targets for photopion production. A cosmic ray that
starts its journey with energy higher than $E_{gzk}$ should loose
rather rapidly the energy in excess of $E_{gzk}$ in the form of
pions. There is great interest in recent observations of cosmic
rays with energies beyond the GZK cutoff~\cite{agasa}~\cite{fly}.
\footnote{This AGASA-data-based ``GZK puzzle'' has been very
important in providing motivation for studies of Planck-scale
departures from Lorentz symmetry. Even if
a future improved understanding of the cosmic-ray spectrum ends up
removing the puzzle, the lessons learned for the study of the
quantum-gravity problem will still be very valuable. An analogous
situation has been rather recently encountered in the
particle-physics literature: the discussion of the
so-called ``centauro events'' led to strong theoretical progress in the
understanding of the possibility of ``misaligned vacua'' in QCD
(see, {\it e.g.}, Ref.~\cite{bjpap}), and this progress on the
theory side remains valuable event though now most authors believe
that ``centauro events'' might have been a mirage }.
It has been
suggested~\cite{bwCR} that perhaps some of the UHE cosmic rays
originate from within our galaxy. This would allow them to evade
the GZK cutoff, since over ``short'' (galactic) distances the
expected energy loss through photopion production is negligible.
But then we should identify within our galaxy some sources that
could accelerate protons to such high energies.

So there are issues of interest for the analysis of acceleration
mechanisms also in the context of cosmic-ray studies, although
of rather different nature with respect to the case of
gamma-ray bursts.

\section{Cosmic-rays as a quantum-spacetime laboratory}
As I stress in my contribution~\cite{gachunt},
the mentioned fact that some cosmic-ray observatories
have reported above-GZK events has also generated
strong interest~\cite{kifu}~\cite{ita}~\cite{gactp}
from the quantum-spacetime research community.
This interest originates from the observation
that spacetime granularity, besides affecting the laws
of particle propagation, can also affect the energetic balance
of particle-physics processes.
The conventional estimate of the GZK cutoff implicitly assumes
that the photopion-production process would occur in an exactly
smooth classical spacetime.
The GZK scale is set by the minimum energy that, in classical spacetime,
is required of a proton to produce a pion in a collision with a photon
of CMB energy.
In some quantum spacetimes this estimate of the minimum energy
is shifted upward by a Planck-scale effect.
Of course if the ``quantum-spacetime GZK scale'' is higher than
estimated classically it would be natural to expect
some cosmic-ray observations that are above the classical GZK scale.

Essentially these quantum-spacetime-inspired studies are
using the cosmic-ray context to probe
a regime of high boosts which is not accessible
in laboratory experiments.
The photopion-production process, $p + \gamma \rightarrow p + \pi$,
is well understood and studied in the laboratory at center-of-mass
energies that are comparable to the ones available in collisions
between a $10^{20}eV$ proton and a CMB photon,
but in our laboratories (when the center-of-mass
energies are so high) we are only able to study the process in frames
that are not highly boosted with respect to the center-of-mass frame.
In the observation of ultra-high-energy cosmic rays
we are instead observing (some consequences of)  the photopion-production
process in a frame which is highly boosted with respect to the
center-of-mass frame, involving indeed an extremely hard proton and
a very soft photon.
The nature of boosts is in one-to-one relation~\cite{gachunt} with the short-distance
structure of spacetime, and therefore it is not surprising
that these studies would be of interest for the quantum-spacetime
community.

I here want to stress that once again the convergence of interest
on a problem (in this case the cosmic-ray-spectrum problem)
by the ``acceleration mechanisms community'' and the ``quantum-spacetime
community'' is accidental.
The new physics associated with spacetime quantization
is not being advocated
as a way to explain the high energies reached by cosmic rays.
Instead the quantum-spacetime
community just uses the experimentally-established fact that some cosmic rays
have huge energies. From a quantum-spacetime perspective it does not matter
how cosmic rays are accelerated to such high energies; it is just
a wonderful opportunity that such high-energy particles are
available for study.

\section{Interplay between the understanding of the emission mechanisms
and the quantum-spacetime studies}
Up to this point I have stressed that the growing number of instances
in which the interests of
the ``acceleration mechanisms community'' and of the ``quantum-spacetime
community'' converge is largely accidental.
This needed to be stressed since readers who have not been following
closely the development of
these fields  might quickly assume that the new physics
of quantum spacetime is being advocated just as a way to device
new acceleration mechanisms, while this is usually not the case.
However, I should also stress that
there are some issues that require clarification from
an acceleration-mechanism perspective and that are quite crucial
for the success of the quantum-spacetime studies.

I want to illustrate this point through a specific example
that is relevant for the gamma-ray-burst studies here mentioned
in Sections~2 and 3.
Basically the quantum-spacetime interest in gamma-ray-burst
observations originates from the fact that the new Planck-scale
laws of particle propagation might attribute different speeds to
photons of different energies.
The fact that
we see some microbursts within a given burst that reach different
energy channels of our
detectors at the same time, within the accuracy available at
our observatories,
allows us to set limits on this energy dependence of the
speed of photons.
The Planck-scale effect could be ``discovered''
if future more sensitive observatories eventually showed
this energy dependence.
But of course the analysis would be severely affected if there
were poorly understood at-the-source correlations between
energy of the photons and time of emission.
As observed recently in Ref.~\cite{piranKARP}
it appears that one can infer such an energy/time-of-emission
correlation from gamma-ray-burst data.
The quantum-spacetime studies will be therefore confronted
with a severe challenge of background/noise removal.
It will be crucial for the quantum-spacetime analysis
to have available a reliable description of the emission
mechanism, which could allow to remove the
undesired at-the-source effect.

\section{Other areas of fundamental physics}
I have so far focused on the fact that some areas of interest from
the acceleration-mechanism perspective are also of interest
for the investigation of certain quantum-spacetime scenarios.
In closing, I want to stress that some of the relevant phenomena
are also relevant for other types of fundamental-physics studies.
Again,
I just illustrate this point through a specific example,
which is relevant for the cosmic-ray studies here mentioned
in Sections~4 and 5.

As mentioned, the most energetic cosmic rays are most likely protons;
however, this does not necessarily imply that they are emitted
as protons. We infer that they are protons on the basis of the nature
of their interactions in the atmosphere, but it is plausible that
the cosmic ray might have started off as some other particle, which then
decays into a proton at a relatively small distance from the Earth.
This would also be another way to describe the observations of
cosmic rays with energies higher than the GZK scale:
it could well be that the cosmic ray is
originally some exotic particle, which does not loose energy through
interactions with CMB photons,
and then this particle decays into a proton (plus other particles)
only at a relatively small distance from the Earth, when the residual
time of travel is not sufficient for substantial energy loss
through interactions with CMB photons.

Various new types of particles, that are independently of interest
from a particle-physics model-building perspective, have been considered
(see, {\it e.g.}, Refs.~\cite{crWIMP1}~\cite{crWIMP2}~\cite{crWIMP3})
in this cosmic-ray context.
Progress in the understanding of the cosmic-ray spectrum could
provide insight on these new particle-physics scenarios.

%
%
%
%
%
%
%
%



\end{document}